\documentstyle[a4,12pt,epsfig]{article}
\begin{document}
\begin{flushright}
DO-TH 00/18;SUSX-TH/00-021;LPTHE/00-46;LA-00-5852;hep-ph/0011395\\
June 2001
\end{flushright}

\vspace{20mm}
\begin{center}
{\Large\bf Dynamics of $O(N)$ chiral
supersymmetry at finite energy density}

\vspace{10mm}

{\bf J. Baacke}$^{(a)}$, 
{\bf D. Cormier}$^{(b)}$, {\bf H. J. de Vega}$^{(c)}$, and 
{\bf K. Heitmann}$^{(d)}$

\vspace{4mm}

{\footnotesize (a) Institut f\"ur Physik, Universit\"at Dortmund, D-44221 Dortmund, 
Germany\\
(b) Centre for Theoretical Physics, Sussex University, Falmer,
Brighton, BN1 9QH, England\\
(c) Laboratoire de Physique Th\'eorique et Hautes Energies,
Universit\'e Pierre et Marie Curie (Paris VI),
Tour 16, 1er. \'etage, 4, Place Jussieu 75252 Paris, Cedex 05, France
 
(d) T-8, Theoretical Division, MS B285, Los Alamos National Laboratory,
Los Alamos, 

New Mexico 87545, U.S.A.}
\\
\vspace{8mm}
{\bf Abstract}
\end{center}
We consider an $O(N)$ version of a massive, interacting, chiral
supersymmetry model solved exactly in the large $N$ limit.  We
demonstrate that the system approaches a stable attractor at high
energy densities, corresponding to a non-perturbative state for which
the relevant field quanta are massless.  The state is one of
spontaneously broken $O(N)$, which, due to the influence of supersymmetry, 
does not become restored at high energies.  Introducing soft
supersymmetry breaking to the Lagrangian results in scalar masses at
the soft breaking scale $m_s$ independent of the mass scale of
supersymmetry $\mu$, with even smaller masses for the fermions.

\newpage

\section{Introduction}

Supersymmetry\cite{bailinlove,sohnius} at finite temperature has a 
number of interesting 
properties\cite{daskaku,boyan,chia,das,DL}.  
Foremost is the appearance of a massless
fermion mode, the so-called Goldstino.  As finite temperature 
corrections are different for particles obeying Bose-Einstein and 
Fermi-Dirac statistics, a finite temperature state is not invariant
under supersymmetry transformations.  Since the transformation parameter
of supersymmetry is a Grassmann variable, the breaking of supersymmetry
therefore implies the existence of a massless Goldstone fermion.

One situation where such properties might be particularly relevant 
is during the reheating stage after inflation.  It has been shown
that light fermions may play an important role in the reheating 
process\cite{BHPfermion}, and so it is important to know whether 
massless or nearly massless fermions might be a requirement in 
supersymmetric models due to the phenomenon of the Goldstone fermion.
However, there is an important distinction between finite temperature
physics and the far from equilibrium, finite energy density situation
relevant to inflation.

Despite the rich activity in out of equilibrium field theory, the case
of supersymmetric models has not yet been properly investigated.  In
this letter, we address a supersymmetric model including all of the
superpartner degrees of freedom, including the fermions.  The model is
one with a global $O(N)$ symmetry which we solve exactly in the limit
of large $N$\cite{largeN}.

Working at large finite energy density, but far from thermal 
equilibrium, we study the dynamics of the model and see that 
the system indeed evolves to an attractor state for which a set 
of fermion modes becomes massless.  We argue that this state is related 
to the spontaneous breaking of the $O(N)$ symmetry and that, therefore,
the fermions are massless because they are superpartners of 
Goldstone bosons.  Somewhat surprisingly, the $O(N)$
symmetry is not restored at high energy as would normally be
expected\cite{boyanovskyjulien,symrest}.  As the energy is increased, 
the contributions from 
the various superpartners to the effective masses of the particle 
modes cancel each other, a result due to supersymmetry.

We stress that although there are $O(N)$ symmetric vacua in this
theory which are degenerate in energy with the $O(N)$ broken vacua, 
at high energies the system invariably evolves to the
$O(N)$ broken vacua for which there are massless fermions.  
In this way, the final vacuum state of the system is predetermined
by the high energy evolution.  The possibility that
there may be selection rules between degenerate vacua resulting
from the early evolution of the universe is a primary result of 
this work.

The model we study contains an $O(N)$ singlet superfield,
characterized by the non-zero vacuum expectation value of the 
scalar component, $\phi(t)$, and an $O(N)$ vector supermultiplet.
The $O(N)$ vector fields have time dependent masses determined 
by an order parameter $m_-^2(t)$ reflecting the internal dynamics
of the system.  If $\phi(t=0)$ is far enough from the supersymmetric
vacua, i.e., if the initial energy density is sufficiently large,
the system evolves in such a way that the order parameter $m_-^2(t)$
asymptotically vanishes, corresponding to zero masses for all of
the fields of the $O(N)$ vector multiplet.  The state is indicative
of the spontaneous breaking of the $O(N)$ symmetry.

We note that while the ground state upon which this
finite energy state is built is supersymmetric, the energy 
is distributed differently among the fermions and bosons 
due to their differing statistics. 
This is analogous to what is known from equilibrium 
studies at finite temperature\cite{daskaku,boyan,chia,das,DL}.
In a cosmological context it is expected that the quanta in the 
highly excited state would become diluted with expansion and the 
universe would eventually find itself approaching the underlying
ground state.

It is also worth mentioning that the state is stable to the 
introduction of small soft supersymmetry breaking to the Lagrangian.  
Such terms explicitly break the supersymmetry, while not introducing
any terms which disrupt the supersymmetric solution to the hierarchy
problem.  Here we find that the qualitative behavior is unchanged
by such terms as the system still evolves to an attractor $O(N)$
broken state.  Furthermore, such terms in fact introduce small 
masses to one set of scalar fields of order the soft breaking 
scale $m_s$, which is taken to be much smaller than the overall 
scale of supersymmetry $\mu$.  The fermions meanwhile gain a mass 
several orders of magnitude smaller.

Although this is only a toy model, the important features of the 
model -- the combination of continuous symmetries leading to 
Goldstone bosons and supersymmetries leading to massless superpartners
-- are completely general.  We also note that in models without 
additional continuous symmetries, such as the ordinary Wess-Zumino
model, the far from equilibrium system is found to evolve toward
a state for which the fermions become massless; this case will be
studied in detail in a future publication\cite{bigsusy}.
The results may be important to our 
understanding of aspects of cosmology, such as supersymmetry 
based inflation models\cite{lythriotto} and electroweak 
baryogenesis\cite{myint,giudice,espinosa,laine}.

\section{The model}

The $O(N)$ extension of the Wess-Zumino model\cite{Wess74} 
consists of a chiral superfield multiplet $S_0=(A_0,B_0;\psi_0;F_0,G_0)$, 
which acts as a singlet under $O(N)$, coupled to $N$ chiral superfields 
$S_i=(A_i,B_i;\psi_i;F_i,G_i)$ with $i = 1\dots N$, transforming 
as a vector under $O(N)$.  Here, $A$ and $F$ are real scalars, $B$ and 
$G$ are pseudo-scalars, and $\psi$ is a Majorana fermion.
The superpotential has the form
\begin{eqnarray}
W(S_0,S_i) &=&  \frac12 M S_0^2
+ \frac{\kappa}{6\sqrt{N}}
S_0^3  \nonumber \\
&+&  \sum_{i=1}^N \frac12 \mu S_i^2 
+ \frac{\lambda}{2\sqrt{N}}S_0 S_i^2   \; .
\label{susyNlagrangian}
\end{eqnarray}

We expand in terms of the component fields and eliminate
the auxiliary fields via their equations of motion.
To allow for a proper large $N$ limit, the expectation value of $A_0$ 
must be of order $\sqrt{N}$.  We therefore set
\begin{equation}
\langle A_0 \rangle = \sqrt{N}\phi \, \, , \; \langle B_0 \rangle = 0
\; .
\end{equation}
The field $B_0$ is taken to have zero expectation value for the sake of
simplicity. We assume that the initial state satisfies the $O(N)$ symmetry 
which requires that $\langle A_i \rangle = \langle B_i \rangle = 0$.
It is convenient to take full advantage of this symmetry 
to define fields $A$,
$B$, and $\psi$ such that $\sum_i A_i A_i = N A^2$, 
$\sum_i B_i B_i = N B^2$, and 
$\sum_i \overline{\psi}_i\psi_i = N \overline{\psi}\psi$.  
The resulting Lagrangian to leading order in $N$ is
\begin{eqnarray}
\frac{{\cal L}}{N} &=& \frac12 \partial_\mu \phi \partial^\mu \phi 
+\frac12 \partial_\mu A \partial^\mu A
+\frac12 \partial_\mu B \partial^\mu B \nonumber \\
&&-\frac12 \phi^2 \left(M + \frac12 \kappa \phi \right)^2
-\frac12 \left(\mu + \lambda \phi\right)^2 \left(A^2 + B^2\right)
\nonumber \\
&&-\frac{\lambda}{2}\left[M\phi + \frac12 \kappa \phi^2 
+ \frac14 \lambda \left(A^2 - B^2\right)\right] \left(A^2 - B^2\right)
\nonumber\\
&&+\frac{i}{2}\bar\psi\gamma^{\mu}\partial_\mu \psi-\frac 1 2 \mu \bar\psi\psi
-\frac{1}{2}\lambda \phi \bar\psi\psi \; .
\label{largeNlagrangian}
\end{eqnarray} 

The system may be completely characterized by 
the equation of motion for the mean field $\phi$
\begin{equation}
\label{mean}
\ddot{\phi} + \frac{M+\kappa \phi}{\lambda}m_-^2 
+ \lambda m_\psi \langle A^2+B^2 \rangle 
+ \frac12 \lambda \langle \overline{\psi}\psi \rangle = 0
\; ,
\end{equation}
and by the time dependent masses of the $\psi$, $A$, and $B$ fields:
\begin{eqnarray}
\label{defmmin}
m_\psi & = & \mu+\lambda\phi \; , \\
\label{lnflmass1}
m_A^2 &=& m_\psi^2 + m_-^2 \; , \\
\label{lnflmass2}
m_B^2 &=& m_\psi^2 - m_-^2 \; ,
\end{eqnarray}
with,
\begin{equation}
m_-^2  \equiv  \lambda\left[M\phi+\frac12 \kappa \phi^2
+ \frac12 \lambda \langle A^2 - B^2\rangle \right] \; . 
\label{m-eqn} 
\end{equation}
The expectation values, $\langle \overline{\psi}\psi \rangle$, 
$\langle A^2 \rangle$, and $\langle B^2 \rangle$ are determined
in the usual way in terms of the non-equilibrium mode functions
for the individual fields.  General details, including the 
renormalization procedure may be found in 
Refs.~\cite{BHPfermion,boyanetal,BHPboson}.
Specific details for this model will be provided in future 
work\cite{bigsusy}.

Note that the appearance of $\langle A^2-B^2 \rangle$ on the right
hand side of the expression for $m_-^2$, Eq.~(\ref{m-eqn}), means
that this expression plays the role of a gap equation which must
be satisfied by the dynamics.  Also note that the sum rule
$m_A^2 + m_B^2 - 2m_\psi^2 = 0$ is automatically satisfied for
all times as required by supersymmetry.

Through use of the equations of motion, it
is straightforward to show that the variation of the Lagrangian
(\ref{largeNlagrangian}) vanishes under supersymmetry transformations
up to a total derivative.  To this order, the Lagrangian is 
completely supersymmetric with $\phi$ acting as a classical 
background field.

We stress that the evolution determined by equation (\ref{mean}) and 
the time dependent masses (\ref{defmmin}) -- (\ref{m-eqn}) exactly 
solves the quantum field theoretical system described by the 
superpotential (\ref{susyNlagrangian})
in the $N \to \infty$ limit to all orders in perturbation theory.  
We have made no further approximations
and the only simplifications, eg., $\langle B_0 \rangle = 
\langle A_i \rangle = \langle B_i \rangle = 0$, correspond to choices of
initial conditions.

\section{Numerical results}

\begin{figure}
\epsfig{file=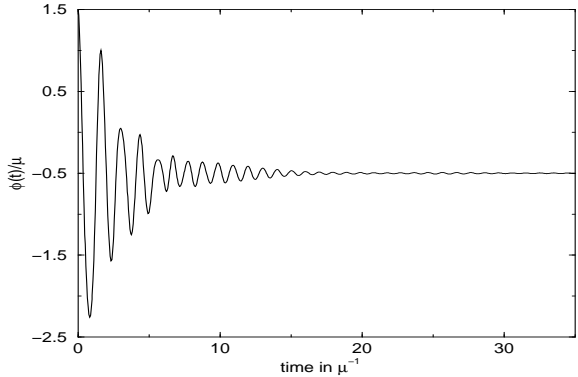,width=3in,height=2.0in}
\caption{Zero mode evolution; the parameters 
are $\mu=1$, $M=4$, $\kappa=1$, $\lambda=2$, and 
$\phi(0)=1.5$.  All masses are scaled by $\mu$ which is arbitrary.}
\label{brokenONphi}
\end{figure}

\begin{figure}
\epsfig{file=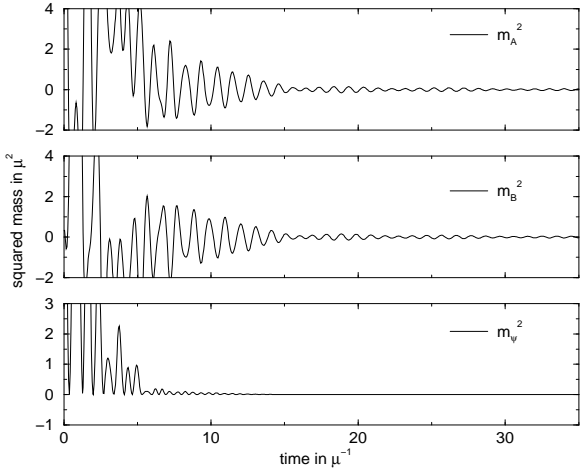,width=3in,height=2.5in}
\caption{The effective field masses squared, $m_A^2$ (top),
$m_B^2$ (middle), and $m_\psi^2$ (bottom) with the same
parameters as in Fig.~\ref{brokenONphi}.  Each effective
mass vanishes at late times.}
\label{brokenONm}
\end{figure}

Finite energy density is imposed via out of equilibrium 
initial conditions for the zero mode $\phi$.
The evolution toward the non-perturbative attractor state is 
depicted in Fig.~\ref{brokenONphi} with the corresponding
masses of the fields in Fig.~\ref{brokenONm}.  We note the 
following characteristics.  The evolution begins with large
oscillations of $\phi$ over the entire classically allowed 
range of evolution.  During this initial period, the field
fluctuations $\langle A^2 \rangle$ and $\langle B^2 \rangle$
grow.  After a relatively short period of time, the mean field 
settles down precisely to the point $\phi = -\mu/\lambda$.  
The result is that the $N$ fermions become massless.

Interestingly, the scalar fluctuations continue to 
grow until a state is formed for which the fields $A$ and $B$ are 
massless as well (Fig.~\ref{brokenONm}).  This configuration remains
completely stable.  We also find that this behavior persists up to
energy densities much higher than any natural scale in the problem,
indicating that there is no symmetry restoration and that this 
represents the generic behavior of the system.  This is made
possible by the cancellation of contributions from 
$\langle A^2 \rangle$ and $\langle B^2 \rangle$ in the gap equation
(\ref{m-eqn}), a direct consequence of supersymmetry, which allows 
these field fluctuations to become
arbitrarily large while not changing their contributions to the
masses of the field quanta.  The growth of these fluctuations
also provides the mechanism for driving the fermion mass to zero,
see Eq.~(\ref{mean}).

In order to understand the supersymmetric field configuration reached
by the non-equilibrium time evolution, it is illuminating to study the 
effective potential for static field configurations  
as a function  $\phi$. It is obtained (see, e.g.
\cite{boyanovskyjulien}) by maximizing 
with respect to $m_-^2$ the potential
\begin{eqnarray}
V(\phi,m_-^2) &=& \left(M\phi+\frac{1}{2}\kappa\phi^2\right)
\frac{m_-^2}{\lambda} -\frac{m_-^4}{2\lambda^2} \nonumber \\
&+& \frac{1}{64\pi^2}\left[g(m_A^2)+g(m_B^2)-2g(m_\psi^2)\right] \; ,
\end{eqnarray}
where  $g(m_i^2)=m_i^4(\ln(m_i^2/m^2)-3/2)$.

\begin{figure}
\epsfig{file=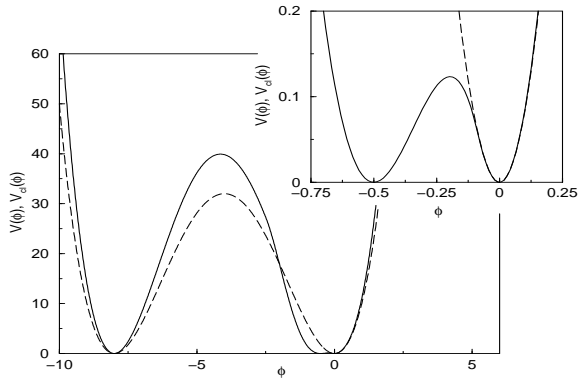,width=3in,height=2in}
\caption{The large $N$ effective potential (solid line), showing
the $O(N)$ symmetric ground states at $\phi=0$ and $\phi=-2M/\kappa=-8$ 
and the spontaneously broken $O(N)$ coexistence phase 
at $\phi=-\mu/\lambda=-1/2$.  The inset frame
magnifies the region around the $O(N)$ broken minimum.
In each case, the tree level potential (dashed line) is shown for 
comparison.  The parameters are as in Fig.~\ref{brokenONphi}.}
\label{effpot}
\end{figure}

The effective potential $V(\phi)$
is plotted in Fig.~\ref{effpot} along with the tree level potential.
We see that the effective potential has a minimum at
$\phi=-\mu/\lambda$, and one finds that at this point all the
masses vanish.  
To understand this new minimum, it is helpful to remember that, 
if one allows for states which break $O(N)$, such that, for example,
$\langle A_1 \rangle = \sigma \neq 0$ then one finds $O(N)$ breaking
minima at $\lambda \phi = -\mu$ and 
$\lambda \sigma = \pm \sqrt{2\mu M - \kappa \mu^3/\lambda}$.  As a result
of the convexity theorem, the exact large $N$ effective potential 
must be flat between these minima as in a Maxwell construction.
Hence, there must be a new minimum in the full effective potential
as shown in the figure.

The state that is reached by the evolution is therefore a 
phase consisting of different spontaneously broken $O(N)$ states
in coexistence.

The values of $\phi$ and the mass parameters obtained at late times 
in the out of equilibrium evolution correspond precisely to such a 
state.  The state itself is a highly excited one 
and is time dependent in a coherent way. 
Furthermore, we find that the system
is attracted towards this configuration once the
energy density is  sufficiently high.

As a side note, we point out that the effective potential in the 
$\phi$ direction is non-convex.  This has to do with the fact that 
$\phi$ acts as a simple classical zero mode whose fluctuations are
completely negligible in the leading order large $N$ limit.

\begin{figure}
\epsfig{file=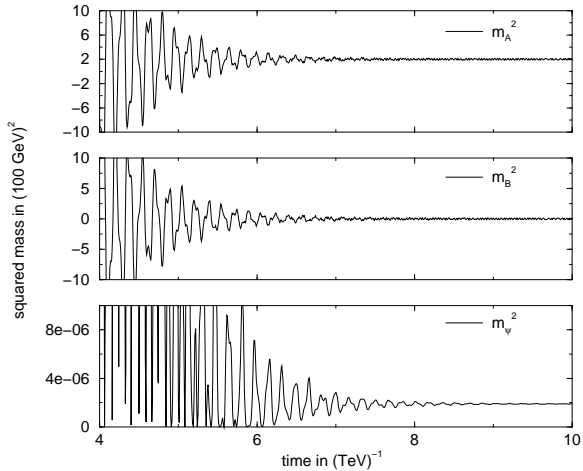,width=3in,height=2.5in}
\caption{The effective field masses squared $m^2_A(t)$, $m^2_B(t)$,
and $m^2_\psi(t)$ in the $O(N)$ broken phase including soft masses 
for the $A$ and $B$.  The parameters are $\mu=10$ TeV, 
$M=40$ TeV, $m_s=100$ GeV, $\kappa=1$, $\lambda=2$, $\phi(0)=15$ TeV.  
The late time values correspond to $m_A=\sqrt{2}m_s=140$ GeV, $m_B=0$, 
and $m_\psi=140$ MeV.}
\label{softm}
\end{figure}

Next, we introduce soft symmetry breaking to the $O(N)$ model via 
a scalar mass $m_s$ for the $A$ and $B$ fields such that 
$m_A^2 = m_\psi^2 + m_-^2 + m_s^2$ and
$m_B^2 = m_\psi^2 - m_-^2 + m_s^2$.    
Fig.~\ref{softm} shows the result for a value $m_s^2/\mu^2 = 10^{-4}$.
We see again that the system reaches the attractor state, 
and the explicit soft symmetry breaking terms provide a mass for 
the $A$ field equal to $\sqrt{2}m_s$, while the $B$ field remains 
massless.  The fermion mass is found to be three orders of magnitude
smaller. 

We make the following conclusions.  First, supersymmetry may play
a very important role in the dynamics of the early universe beyond
ordinary model building.  The requirement of massless fermions 
appearing in the spectrum, in particular, may be important to 
inflationary and reheating dynamics and could also play a significant 
role in baryogenesis.  Furthermore, the constraints of supersymmetry 
can lead to a preferential choice between multiple degenerate vacua
through the existence of attractor states at finite energy density.
Such issues are deserving of further study.

\section*{Acknowledgements} 
The authors would like to thank Dan Boyanovsky for directing their 
interest toward out of equilibrium supersymmetry.

\end{document}